\documentclass[aps,pra,showpacs,twocolumn,superscriptaddress]{revtex4-1}
\usepackage{amsmath}
\usepackage{amssymb}
\usepackage{graphicx}
\usepackage{epstopdf}
\usepackage{times,txfonts}
\usepackage{easybmat}
\usepackage{mathtools}

\usepackage[bookmarks=false]{hyperref}
\hypersetup{colorlinks=true,citecolor=blue,linkcolor=blue,urlcolor=blue,pdfstartview=FitH,bookmarksopen=true}
\usepackage{color,soul}

\begin{document}
\title{Measurement-based universal blind quantum computation with minor resources}
\author{Xiaoqian Zhang}
\email{zhangxq67@mail.sysu.edu.cn}

\affiliation{School of Physics and State Key Laboratory of Optoelectronic Materials and Technologies, Sun Yat-sen University, Guangzhou 510000, China}
\date{\today}

\begin{abstract}
  Blind quantum computation (BQC) enables a client with less quantum computational ability to delegate her quantum computation to a server with strong quantum computational power while preserving the client's privacy. Generally, many-qubit entangled states are often used to complete BQC tasks. But for a large-scale entangled state, it is difficult to be described since its Hilbert space dimension is increasing exponentially. Furthermore, the number of entangled qubits is limited in experiment of existing works. To tackle this problem, in this paper we propose a universal BQC protocol based on measurement with minor resources, where the trap technology is adopted to verify correctness of the server's measurement outcomes during computation and testing process. In our model there are two participants, a client who prepares initial single-qubit states and a server that performs universal quantum computation. The client is almost classical since she does not require any quantum computational power, quantum memory. To realize the client's universal BQC, we construct an $m\times n$ latticed state composed of six-qubit cluster states and eight-qubit cluster states, which needs less qubits than the brickwork state. Finally, we analyze and prove the blindness, correctness, universality and verifiability of our proposed BQC protocol.
\end{abstract}

\maketitle

\section{Introduction}

Quantum computation has already been widely studied by different styles \cite{1Deut,2Deutsch73,3Robert,4Rich}. The quantum logic network \cite{2Deutsch73} can be used to establish a relationship between quantum physics and quantum information processing. The quantum technology is improved continuously, which makes it possible for the first generation quantum computers to come out. Quantum computers can only be possessed by companies and governments because of their expensive prices for average persons. However, quantum computing will become essential for most people in the future. When people want to perform quantum computing, quantum computers or simulators can be used as quantum cloud platforms to satisfy such requirements. In this case, the client's quantum computing can be delegated to these quantum cloud platforms called servers. This delegation will bring a key problem, that is how to guarantee the client's quantum computing privacy. To be specific, servers only obtain the information that the client tells, but cannot get anything else. To solve the problem better, blind quantum computation (BQC) technology is adopted.

For this problem, numerous blind quantum computation protocols are proposed \cite{4Childs2005,5Fisher2014,6Broa15,7Delgado2015,8Elham,9Morimae,10Broadbent09,11Morimae2015,12Morimae2012,
13Tomoyuki,14Barz,15Sueki,16Mantri2013,17Giovanne2013,18Mantri,19Morimae13,20Li14,21He,4Sheng15,15Takeuchi2016}. As we know, blind quantum computation is a new secure quantum computing, in which a client with less quantum technologies outsources her computation to a server with a fully-fledged quantum computers. In the process, the client's quantum abilities are not sufficient for universal quantum computation and any of her secret information will not be leaked to servers. Broadbent \emph{et al.} \cite{10Broadbent09} proposed an universal blind quantum computation based on a brickwork state (which is called BFK protocol), which allows a client to delegate quantum computation to a server while remaining the client's inputs, outputs and computation perfectly private. In their protocol, the client is able to prepare single qubits randomly chosen from a finite set $\frac{1}{\sqrt{2}}(|0\rangle+e^{i\theta}|1\rangle)(\theta\in\{0, \frac{\pi}{4}, \frac{2\pi}{4}, \ldots, \frac{7\pi}{4}\}$ and the server has the ability to control quantum computational resources. Moreover, a fault-tolerant authentication protocol was given to verify an interfering server. Barz \emph{et al.} \cite{14Barz} made an experiment to demonstrate blind quantum computing, where the client had the abilities to prepare and transmit individual photonic qubits keeping input data, algorithms, and output data private. Naturally, measurement-based multi-server BQC protocols with Bell states \cite{19Morimae13,20Li14} were proposed, which were degraded to single-server performing BFK protocol \cite{10Broadbent09}. Besides blind brickwork state, BQC protocols based on blind topological states \cite{12Morimae2012} and Affleck-Kennedy-LiebTasaki (AKLT) \cite{11Morimae2015} were also studied respectively.

However, servers are almost semi-trusted in BQC, which makes it an urgent problem to detect the correctness of servers' computing results. Aiming at such a problem, many methods \cite{7Morimae2014,6Hayashi2015,Gheor15,Fit15,Moe16,Anne17,Alex16} can be employed to  realize the verifiability such as verifying quantum inputs \cite{Moe16} and quantum computing \cite{7Morimae2014}. As for other aspects, many scholars started to consider solving practical questions by using BQC technologies \cite{22Jos,30Huang,13Sun15,25Hua,31Mar}. For instance, Huang \emph{et al.} \cite{30Huang} implemented an experimental BQC protocol to factorize the integer 15 in which the classical client can interact with two entangled quantum servers. Recently, Fitzsimons \cite{22Jos} analyzed and summarized some important BQC protocols in terms of security, state preparation and so on.

It is crucial for quantum computers to prepare entangled states \cite{10Broadbent09,Horodecki09} in large scales of space-separated or individual-controllable quantum systems. For example, one of the genuine entangled states---the brickwork state \cite{10Broadbent09}---was constructed in theory to realize universal blind quantum computation. The large-scale quantum entangled states can be viewed as vital resources in quantum field such as quantum nonlocality \cite{Bandyopa11}, quantum computing \cite{Rausse01} and quantum simulation \cite{Seth96}. Concretely, in terms of realizing quantum parallel computing, a great amount of quantum entanglement makes quantum computers and simulators superior classical computers. Meanwhile, there are some important progress in experiment to prepare multi-qubit entangled states recently. For a trapped-ion system, the number of qubits in an entangled state \cite{Monz11} reaches to $14$ in 2011, while the number merely increases to $20$ deterministically implemented by Friis \emph{et al.} \cite{Friis18} in 2018. In addition, the number of entangled qubits is only $10$ both in superconducting \cite{Song17} and photonic system \cite{Lin16}. However, the Hilbert space dimension will increase doubly when a qubit is added into an experimental system, which becomes a significant challenge to describe the new large entangled state.

In \cite{10Broadbent09}, universal gates H, T, CNOT can be realized by ten-qubit cluster states respectively. To lessen the number of qubits, we propose a measurement-based universal BQC (MUBQC) protocol with a minor resource called latticed state. In this article, there are two participants, a client Alice and a server Bob. In the process of blind quantum computation, we suppose that a client Alice prepares trustworthy initial single-qubit states $|\pm_{\theta}\rangle$, $|0\rangle$ and $|1\rangle$ and a server Bob performs universal quantum computation. In the verifiable process, we assume that the sever Bob is a polynomial time quantum prover and the client Alice is a polynomial time classical verifier. Similar to the Ref.\cite{Moe16}, we assume that a decision problem L needs to be solved by Alice in our protocol. Usually for any instance $x$, if $x \in L$, the acceptance probability is larger than 2/3, and if $x\notin L$, the acceptance probability is smaller than 1/3. The latticed state is composed of two classes of cluster states: six-qubit cluster states mainly realizing gates S, Z, T, X, Y, I and eight-qubit cluster states chiefly realizing gates H, CNOT. Therefore, our proposed latticed state with less qubits is possible to be prepared in the laboratory. Furthermore, we respectively prove the blindness, correctness, universality and verifiability of our protocol. These factors are often considered in other BQC protocols. Notice that the verifiability means to verify the correctness of Bob's measurement outcomes in computing and testing process, which is achieved by trap technology in measurement. The proof technology of verifiability refers to the work in \cite{Moe16}. The employed encrypted method is from the BFK protocol in \cite{10Broadbent09}.

The rest of paper is organized as follows. The blind quantum computation protocol is presented in Sec. \ref{sec:jud} as well as analyzing and proving the blindness, correctness, universality and verifiability. Finally, the conclusions are shown in Sec. \ref{sec:con}.

\section{Measurement-based universal BQC protocol}
\label{sec:jud}
In this section, we construct the latticed state for the first time and design our measurement-based universal BQC (MUBQC) protocol. And then we give out analyses and proofs with respect to the blindness, correctness, universality and verifiability of our MUBQC protocol.

\begin{figure}[!htp]
  \centering
  \includegraphics[scale=0.45]{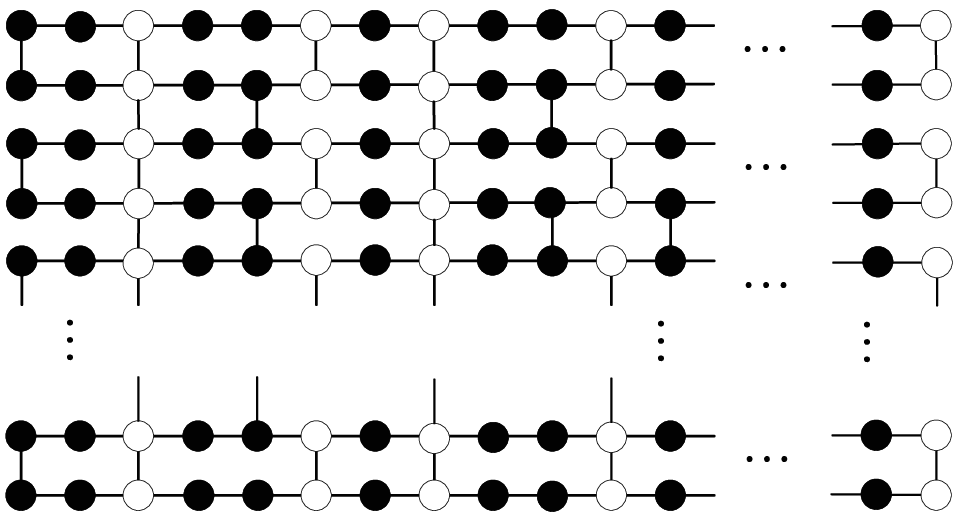}
  \caption{\ The structure of the latticed state $|LA\rangle$.}\label{fig1}
\end{figure}

\emph{Definition of the latticed state.}---An $m \times n$ dimensional latticed entangled state $|LA\rangle$ is constructed as follows (see Fig. \ref{fig1}). Here, we set that $m$ represents the total number of horizontal rows and $n$ represents the total number of vertical columns. To express  conveniently, we suppose $N=mn$ in the follow-up description.\\
1. All original qubits are in states $|\pm_{\kappa_t}\rangle=\frac{1}{\sqrt{2}}(|0\rangle\pm e^{i\kappa_t}|1\rangle)$, where $\kappa_t=0, \frac{\pi}{4}, \ldots, \frac{7\pi}{4}$.\\
2. We label physical qubits with indices $x$ and $y$. Here $x$ represents the $x^{th}$ row and $y$ represents the $y^{th}$ column.\\
3. For each column, apply operations controlled-Z (CZ) on qubits $(x,y)$ and $(x,y+1)$ where $1\leqslant x\leqslant m, 1\leqslant y\leqslant n$.\\
4. For odd rows $x$ and columns $y\equiv 1\ (mod \ 5)$, apply operations CZ on qubits $(x,y)$ and $(x+1,y)$, $(x,y+2)$ and $(x+1,y+2)$.\\
5. For even rows $x$ and columns $y\equiv 3\ (mod\ 5)$, apply operations CZ on qubits $(x,y)$ and $(x+1,y)$, $(x,y+2)$ and $(x+1,y+2)$.\\
6. The white circles denotes the computational outputs of previous cluster states and the inputs of latter cluster states at the same time, while the black circles denote auxiliary qubits for realizing quantum computing.

Specifically, the latticed state $|LA\rangle$ in Fig. \ref{fig1} can be decomposed into six-qubit cluster states and eight-qubit cluster states (respectively Fig. \ref{fig3} and Fig. \ref{fig4} in Appendix A), in which six-qubit cluster states mainly realize gates S, T, X, Y, Z, I and eight-qubit cluster states mainly realize gates H, CNOT. In fact, every eight-qubit cluster state can also be used to achieve gates S, T, X, Y, Z, I, while there is undesirable operations H in Fig. \ref{fig5} (See Appendix A). Therefore, to obtain S, T, X, Y, Z, I, we prefer to use six-qubit cluster states than eight-qubit cluster states for efficiency improvement. For CNOT gate, we notice that it needs correction gates H and $R_z(\textnormal{-}\frac{\pi}{2})$. If the cluster states do not contain quantum outputs, the correction operations $R_z(\textnormal{-}\frac{\pi}{2})$ will be naturally absorbed since Alice can ask Bob to perform the projective measurements $|\pm_{\eta_t-\frac{\pi}{2}}\rangle\Leftrightarrow R_z(\textnormal{-}\frac{\pi}{2})|\pm_{\eta_t}\rangle=\frac{1}{\sqrt{2}}(e^{\frac{i\pi}{4}}|0\rangle\pm e^{i(\eta_t-\frac{\pi}{4})}|1\rangle)=\frac{e^{\frac{i\pi}{4}}}{\sqrt{2}}[|0\rangle\pm e^{i(\eta_t-\frac{\pi}{2})}|1\rangle]$ (See Appendix A). Note that, ${\eta_t}$ comes out in Step 4 of our protocol. Otherwise, the correction operations $R_z(\textnormal{-}\frac{\pi}{2})$ will be performed on useful outputs of qubits and randomly traps to hiding gate CNOT since none of outputs needs to be measured.

\begin{figure}[!htp]
  \centering
  \includegraphics[scale=0.7]{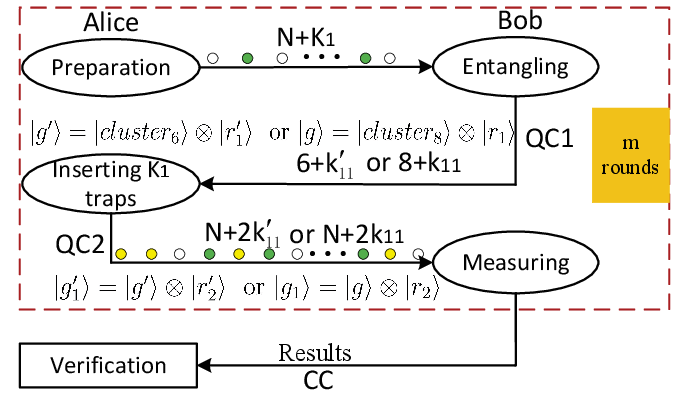}
  \caption{\ (Color online) Schematic diagram of MUBQC protocol, where the green circles and yellow circles are trap qubits $|R_1\rangle$ and $|R_2\rangle$, respectively. $|R_1\rangle$ consists of $|r'_1\rangle$ and $|r_1\rangle$, and $|R_2\rangle$ consists of  $|r'_2\rangle$ and $|r_2\rangle$. QC1 and QC2 represents quantum channels between Alice and Bob. CC is the classical channel.}\label{fig2}
\end{figure}

\emph{MUBQC Protocol.}---The principle of measurement-based quantum computation is presented in Refs. \cite{10Broadbent09,Danos07,Jozsa05} detailedly. In our design, MUBQC protocol can be realized by measuring the latticed state, shown in Fig. \ref{fig2}. From Fig. \ref{fig2}, Alice prepares and distributes enough single-qubit states, while the server Bob does the entanglement and measurements. Notice that, for hiding CNOT, Alice randomly asks Bob to perform operations H on computing qubits $|LA\rangle$ or trap qubits from $|R_1\rangle$ and $|R_2\rangle$. Therefore, Bob can not distinguish these operators performed on $|LA\rangle$ or $|R_1\rangle$ or $|R_2\rangle$ in the stage of Bob's measurement.

Our protocol runs as follows:

1) Alice prepares N single-qubit states $|\pm_{\kappa_t}\rangle=\frac{1}{\sqrt{2}}(|0\rangle\pm e^{i\kappa_t}|1\rangle)$ ($\kappa_t\in \{0, \frac{\pi}{4}, \cdots, \frac{7\pi}{4}\}$) and $K_1$ single-qubit states $|R_1\rangle=\{|0\rangle$, $|1\rangle\}$, where $K_1=2m_1\times n$ ($m_1$ is an integer and $1\leqslant m_1\leqslant \frac{m}{2}$). Alice sends these qubits to Bob. Here, trap qubits $|R_1\rangle$ are randomly attached to the latticed state with certain rules such as Fig. 4 in Appendix A, but from Bob's view, the combination structure of latticed state and trap qubits $|R_1\rangle$ cannot be distinguished from the original latticed state.

In the step 2, we consider two cases.

\begin{figure}[!htp]
  \centering
  \includegraphics[scale=0.5]{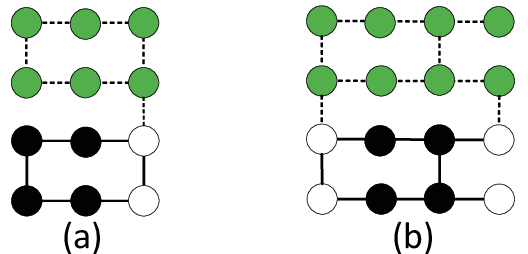}
  \caption{\ (Color online) $(a)$ A six-qubit cluster state with trap qubits, i.e. $|g'\rangle$. $(b)$ A eight-qubit cluster state with trap qubits, i.e. $|g\rangle$.}\label{fig13}
\end{figure}

2A) If Alice wants to realize a single-qubit gate S or T or X or Z or Y or I, Bob performs $CZ$ gate to get state $|g'\rangle=|cluster_6\rangle\otimes |r'_1\rangle$ according to Alice's orders. The $|g'\rangle$ state is shown in Fig. \ref{fig13}(a), where $|cluster_6\rangle$ is used to compute and $|r'_1\rangle=(|0\rangle\otimes |1\rangle)^{\otimes k'_{11}}$ is applied to test the correctness of the $|cluster_6\rangle$ state. Bob returns $|g'\rangle$ to Alice, Alice generates $k'_{11}$ trap qubits $|r'_2\rangle=|\pm_\varphi\rangle$ ($\varphi\in\{0, \frac{\pi}{4}, \ldots, \frac{7\pi}{4}\}$) and randomly inserts them into the sequence containing $(6+k'_{11})$ qubits, in which $k'_{11}\leqslant 6$. Then Alice sends all qubits $|g'_1\rangle$ to Bob again and Bob performs measurements. The output qubits are entangled with other qubits to construct next cluster state. Because of the existence of trap qubits, the useful gates can be concealed.

2B) If Alice wants to realize a gate H or CNOT, Bob performs $CZ$ gate to get state $|g\rangle=|cluster_8\rangle\otimes |r_1\rangle$ according to Alice's orders. The $|g\rangle$ state is shown in Fig. \ref{fig13}(b), where $|cluster_8\rangle$ is used to compute and $|r_1\rangle=(|0\rangle\otimes |1\rangle)^{\otimes k_{11}}$ is applied to test the correctness of the $|cluster_8\rangle$ state. Then Bob returns $|g\rangle$ to Alice, Alice generates $k_{11}$ trap qubits $|r_2\rangle=|\pm_\varphi\rangle$ ($\varphi\in\{0, \frac{\pi}{4}, \ldots, \frac{7\pi}{4}\}$) and randomly inserts them into the sequence containing $(8+k_{11})$ qubits, in which $k_{11}\leqslant 8$. Then Alice sends all qubits $|g_1\rangle$ to Bob again. After measurement, Bob obtains the useful output operated by ($H\otimes I$)CNOT or $H\otimes H$. Then Alice immediately asks Bob to perform ($H\otimes I$) gate and get a CNOT gate or H gate. The output qubits are entangled with other qubits to construct next cluster state. To hide CNOT and H, Alice randomly asks that Bob performs gate H on trap qubits $|0\rangle$, $|1\rangle$, $|\pm_\varphi\rangle$. Note here, $|r'_1\rangle$ and $|r_1\rangle$ belong to $|R_1\rangle$, while $|r'_2\rangle$ and $|r_2\rangle$ belong to $|R_2\rangle$.

Bob repeats the step 2A) or 2B) until all measurements are completed, after that the graph state $|G\rangle$ (See Fig. \ref{fig6}) is constructed spontaneously. In Fig. \ref{fig6}, these trap qubits can be randomly attached to the $|LA\rangle$ state as long as they keep the structural consistency and do not affect the efficient computing.

The qubits from $|r_1\rangle$ and $|r'_1\rangle$ are not entangled, $|custer_6\rangle$ and $|r'_1\rangle$, $|custer_8\rangle$ and $|r_1\rangle$ are not entangled with each other, while qubits in $|custer_8\rangle$ and $|custer_6\rangle$ are entangled with each other. Although we use unit cluster states to realize blind quantum computation, the verifiability process will not be affected since we analyze the whole computation protocol.

It is vital to fix the number of qubits from $|LA\rangle$ and traps $|R_1\rangle$, since too many traps will affect the computational efficiency while too few traps will reduce the probability of checking Bob's deception. In our MUBQC protocol, the number of qubits from $|LA\rangle$ and traps $|R_1\rangle$ is set N and $K_1$ respectively ($N\geqslant K_1$), which exists a tradeoff between the computational efficiency and the probability of checking Bob's deception.

3) For the $t^{th}$ qubit, Alice computes measurement angles $\eta_t=\theta'_t+\kappa_t+r\pi$, where $r\in\{0,1\}$ and $\theta'_t=(-1)^{s_t^X}\theta_t+s^Z_t\pi$. To be specific, the actual measurement angle $\theta'_t$ is a modification of $\theta_t$ that depends on previous measurement outcomes. $\theta_t$ is the specified measurement angle for each qubit. $s^{Z(X)}_t$ is the parity of all measurement outcomes for qubits for Z(X) measurements. We define that the measurement results in the first row and the first column are zero \cite{10Broadbent09}. The measurement angles $\eta_t$ and $\varphi$ belong to the same set $\{0, \frac{\pi}{4}, \cdots, \frac{7\pi}{4}\}$ so that they cannot be distinguished from Bob's view. Alice sends relevant measurement angles $\eta_t$ and $\varphi$ to Bob, where measurement outcomes are always labelled 0 or 1.

4) Bob measures all qubits and returns these results to Alice, where the positions of trap qubits are unknown to Bob. After receiving results from Bob, Alice will performs the following three processes with a certain probability.

5) With probability $q$ ($0\leqslant q\leqslant1$), if the computation result is acceptable after directly abandoning all traps $|R_1\rangle$ and $|R_2\rangle$, the probability of Alice accepting the results of $|LA\rangle$ is at least $\frac{3}{4}$. (If Bob is malicious to randomly prepare a fake graph state, the original states $|\pm_{\kappa_t}\rangle$ are randomly changed into $|0\rangle, |1\rangle, |\pm_{\kappa'_t}\rangle$. Thus, the probability that Alice obtains correct results is $\frac{3}{4}$ while the probability is larger than $\frac{3}{4}$ for an honest Bob). Otherwise, the probability of Alice accepting is at most $\frac{1}{4}$ (If Bob is malicious, the probability that Alice accepts false measurement results is $\frac{1}{4}$ while the probability is less than $\frac{1}{4}$ for an honest Bob). According to the value $r$, Alice determines whether the result is flipped or not when Alice accepts these results.

With probability $\frac{1-q}{2}$, Alice tests the results $|R_1\rangle$ to detect the correctness of the latticed state. If the results returned by Bob are coincide with the values predicted from the outcomes in the original $|R_1\rangle$, then the test is passed.

With probability $\frac{1-q}{2}$, Alice tests the results of $|R_2\rangle$ to check the correctness of measurement results. If the results returned by Bob are coincide with the values predicted from the outcomes in the original $|R_2\rangle$, then the test is passed.
\\

In our protocol, if Bob is honest, Alice will realize her computing successfully. However, if Bob is malicious, he can not get anything about Alice's privacy since Alice can check out the malicious behaviour and abort the protocol.

Notice that, we can ensure that the structure of trap qubits are not distinguished from the original latticed state in Bob's side. In fact, traps in $|R_1\rangle$ do not affect the computation because there is no entanglement not only among qubits in $|R_1\rangle$ but also between $|R_1\rangle$ and $|LA\rangle$. In Fig. \ref{fig6} (Appendix A), we show the structure of the graph state $|G\rangle$ as an example.

\emph{Analyses and proofs}---Here, we will give the analyses and proofs of correctness, blindness, universality and verifiability in detail.

\emph{Theorem 1 (Correctness).} If Alice and Bob follow the steps of our MUBQC protocol, these outcomes will be correct.

\emph{Proof:}
1) In Fig. 4, suppose operations I, S, T, X, Y, Z are performed on the above qubit, and then I is performed on the below qubit. For gates I, S, Z, T, H, these circuits are simple and we directly obtain the Eq.(1), but the simplification process of gates X, Y and CNOT are relatively complicated, as shown in Figs. \ref{fig10}, \ref{fig11} and \ref{fig12} (See Appendix B).
\begin{eqnarray}
\begin{array}{l}
\displaystyle I=HR_z(0)HR_z(0),\qquad S=e^{\frac{i\pi}{4}}H R_z(0)HR_z(\frac{\pi}{2}),\\
\displaystyle T=e^{\frac{i\pi}{8}}HR_z(0)H R_z(\frac{\pi}{4}),\ Z=e^{\frac{i\pi}{2}}HR_z(0)H R_z(\pi),\\
\displaystyle H=HR_z(0)HR_z(0)H R_z(0).
\end{array}
\end{eqnarray}
where $R_z(0)=I$, $(R_z(\theta)\otimes I)CZ=CZ(R_z(\theta)\otimes I)$, $HR_z(\theta)H=R_x(\theta)$. Therefore, the correctness is proved.$\square$

\emph{Theorem 2 (blindness of the latticed state).} The dimension of the latticed state in our MUBQC protocol is private. The positions of six-qubit cluster states and eight-qubit cluster states may leak.

\emph{Proof:}
In our protocol, the graph state prepared by Bob are composed of traps and computational qubits, so it is obviously that the dimension known to Bob is larger than the dimension of the latticed state. That is to say, the dimension of latticed state keeps privacy to Bob except the dimension of brickwork state \cite{10Broadbent09}. If Bob is very careful, he will find that the positions of six-qubit cluster states and eight-qubit cluster states. However Bob can only know this at most since all measurement angles are encrypted and there exists the confusion of traps. Therefore, our construction accords with the blindness property of the latticed state. $\square$

\emph{Theorem 3 (blindness of quantum inputs).} The quantum inputs are $|\pm_{\vartheta_j}\rangle$ ($\vartheta_j=0, \frac{\pi}{4}, \ldots, \frac{7\pi}{4}$) and $|0\rangle, |1\rangle$ which are unknown to Bob.

\emph{Proof:}
We can see that the density matrix is independent of $|\pm_{\vartheta_j}\rangle$ ($\vartheta_j=0, \frac{\pi}{4}, \ldots, \frac{7\pi}{4}$), $|0\rangle$ and $|1\rangle$ as follows.

\begin{eqnarray}
\begin{array}{l}
\displaystyle \frac{1}{18}[\sum\nolimits_{\vartheta_j}[|+_{\vartheta_j}\rangle\langle+_{\vartheta_j}|+|-_{\vartheta_j}\rangle\langle-_{\vartheta_j}|+|0\rangle\langle0|+|1\rangle\langle1|]\\
\displaystyle =\frac{1}{18}[|+\rangle\langle+|+|+_{\frac{\pi}{4}}\rangle\langle+_{\frac{\pi}{4}}|+|+_{\frac{\pi}{2}}\rangle\langle+_{\frac{\pi}{2}}|+|+_{\frac{3\pi}{4}}\rangle\langle+_{\frac{3\pi}{4}}|\\
\displaystyle +|+_{\pi}\rangle\langle+_{\pi}|+|+_{\frac{5\pi}{4}}\rangle\langle+_{\frac{5\pi}{4}}|+|+_{\frac{3\pi}{2}}\rangle\langle+_{\frac{3\pi}{2}}|+|+_{\frac{7\pi}{4}}\rangle\langle+_{\frac{7\pi}{4}}|\\
\displaystyle +|-\rangle\langle-|+|-_{\frac{\pi}{4}}\rangle\langle-_{\frac{\pi}{4}}|+|-_{\frac{\pi}{2}}\rangle\langle-_{\frac{\pi}{2}}|+|-_{\frac{3\pi}{4}}\rangle\langle-_{\frac{3\pi}{4}}|\\
\displaystyle +|-_{\pi}\rangle\langle-_{\pi}|+|-_{\frac{5\pi}{4}}\rangle\langle-_{\frac{5\pi}{4}}|+|-_{\frac{3\pi}{2}}\rangle\langle-_{\frac{3\pi}{2}}|+|-_{\frac{7\pi}{4}}\rangle\langle-_{\frac{7\pi}{4}}|\\
\displaystyle +|0\rangle\langle0|+|1\rangle\langle1|]=\frac{1}{2}I.
\end{array}
\end{eqnarray}
From Eq.(2), Bob cannot obtain anything about the state from $\{|\pm_{\vartheta_j}\rangle, |0\rangle, |1\rangle\}$ since Alice has utilized the depolarizing channel. Even if Bob destroys or tampers the states by some way, he can learn nothing about them. Therefore, we have proved the blindness of quantum inputs. $\square$

\emph{Theorem 4 (blindness of algorithms and outputs).} The blindness of quantum algorithms and classical outputs can be proved by Bayes' theorem.

1) The conditional probability distribution of Bob obtaining computational angles is equal to its priori probability distribution, even if Bob knows all classical information and all measurement results of any positive-operator valued measures (POVMs) at any stage of the protocol.

2) All classical outputs are one-time pad to Bob.

\emph{Proof:} Refer to \cite{11Morimae2015,12Morimae2012}.$\square$

\emph{Theorem 5 (Universality).} The universal quantum computing can be realized by a standard universal gates set H, T, CNOT \cite{2000MAN}.

\emph{Proof:} As we see that, in Fig. \ref{fig3}, quantum gates S, Z, T, X, Y, I can be realized by the help of six-qubit cluster states. Gates H, CNOT can be realized by eight-qubit cluster states in Fig. \ref{fig4}. In Fig. \ref{fig1}, the latticed state contains six-qubit cluster states and eight-qubit cluster states such that all gates can be realized by the combination of gates H, T and CNOT. $\square$

\emph{Theorem 6 (Verifiability).} If Bob is honest, Alice can obtain the correct results. However, if Bob is malicious, he returns fake results.
In measurement-based quantum computation model \cite{McK16,Alex16}, the interactive proof is performed between the server Bob who is a polynomial time quantum prover and the client Alice who is a polynomial time classical verifier. We prove the two items completeness and soundness as follows, where language L belongs to BQP.

1) (Completeness) If $x\in L$, the probability that Alice accepts Bob is at least $\frac{2}{3}$.

2) (Soundness) If $x\notin L$, the probability that Alice accepts Bob is no more than $\frac{1}{3}$.

\emph{Proof:}
Firstly, we prove the completeness as follows. If $x\in L$, honest Bob measures the correct state $|G_1\rangle$ such that Alice obtains the correct results, and the probability of passing the tests are $1$. Therefore, the acceptance probability $P$ is
\begin{eqnarray*}
\begin{array}{l}
\displaystyle P\geqslant (3q)/4+\frac{1-q}{2}\cdot1+\frac{1-q}{2}\cdot1\\
\displaystyle \quad>(2q)/3+\frac{1-q}{2}\cdot1+\frac{1-q}{2}\cdot1\equiv \zeta.
\end{array}
\end{eqnarray*}
where $0\leqslant q\leqslant 1$. Then we have
$$\zeta\equiv (2q)/3+\frac{1-q}{2}+\frac{1-q}{2}=\frac{2}{3}q+(1-q)=1-\frac{1}{3}q\geqslant\frac{2}{3}.$$
Therefore, we prove the completeness.

Next, the soundness is considered. Let $x\notin L$, Bob might be malicious to measure any $(N+2K_1)$-qubit state $|G'_1\rangle$. Suppose $\epsilon\geqslant\frac{2}{3(1-q)}$, we can obtain the acceptance probability $P$ by the following cases. $P_1$ and $P_2$ respectively represent the probability of passing tests in traps $|R_1\rangle$ and $|R_2\rangle$.

1) If $P_1< 1-\epsilon$ and $P_2<1-\epsilon$, then
\begin{eqnarray*}
\begin{array}{l}
\displaystyle P\leqslant q/4+\frac{1-q}{2}(1-\epsilon)+\frac{1-q}{2}(1-\epsilon)\\
\displaystyle \quad<q+\frac{1-q}{2}(1-\epsilon)+\frac{1-q}{2}(1-\epsilon)\equiv\xi_1.
\end{array}
\end{eqnarray*}
Thus we get
\begin{eqnarray*}
\xi_1\equiv q+\frac{1-q}{2}(1-\epsilon)+\frac{1-q}{2}(1-\epsilon)=1-(1-q)\epsilon.
\end{eqnarray*}

2) If only one of tests passes, that is one of $P_1$ and $P_2$ is at least $1-\varepsilon$, then
\begin{eqnarray*}
\begin{array}{l}
\displaystyle P\leqslant q/4+\frac{1-q}{2}\cdot1+\frac{1-q}{2}(1-\epsilon)\\
\displaystyle \quad<q/3+\frac{1-q}{2}\cdot1+\frac{1-q}{2}(1-\epsilon)\equiv\xi_2.
\end{array}
\end{eqnarray*}
Hence, we obtain
\begin{eqnarray*}
\xi_2\equiv q/3+\frac{1-q}{2}+\frac{1-q}{2}(1-\epsilon)=1-\frac{\epsilon}{2}+(\frac{\epsilon}{2}-\frac{2}{3})q.
\end{eqnarray*}

3) If $P_1\geqslant 1-\epsilon$ and $P_2\geqslant1-\epsilon$, then
\begin{eqnarray*}
\begin{array}{l}
\displaystyle P\leqslant (1/4+2\sqrt{\epsilon})q+\frac{1-q}{2}\cdot1+\frac{1-q}{2}\cdot1\\
\displaystyle \quad<(2/3+2\sqrt{\epsilon})q+\frac{1-q}{2}+\frac{1-q}{2}\equiv\xi_3.
\end{array}
\end{eqnarray*}
So, we have
\begin{eqnarray*}
\xi_3\equiv (2/3+2\sqrt{\epsilon})q+\frac{1-q}{2}+\frac{1-q}{2}=1-(\frac{1}{3}-2\sqrt{\epsilon})q.
\end{eqnarray*}

Suppose $\xi_1\geqslant\xi_2$ and $\xi_1\geqslant\xi_3$, we can get two inequalities $(1)$ $\xi_1-\xi_2\geqslant0$ and $(2)$ $\xi_1-\xi_3\geqslant0$ accordingly. By plugging $\xi_1$ and $\xi_2$ in inequality $(1)$, we have
\begin{eqnarray*}
\begin{array}{l}
\displaystyle \xi_1-\xi_2=1-(1-q)\epsilon-1+\frac{\epsilon}{2}-(\frac{\epsilon}{2}-\frac{2}{3})q\\
\displaystyle \qquad\quad \ =-\frac{\epsilon}{2}+(\frac{2}{3}+\frac{\epsilon}{2})q\geqslant0,
\end{array}
\end{eqnarray*}
and it is straightforward that $q\geqslant\frac{\frac{\epsilon}{2}}{\frac{2}{3}+\frac{\epsilon}{2}}=\frac{3\epsilon}{4+3\epsilon}.$ Similarly, by plugging $\xi_1$ and $\xi_3$ in inequality $(2)$, we have
\begin{eqnarray*}
\begin{array}{l}
\displaystyle \xi_1-\xi_3=1-(1-q)\epsilon-1+(\frac{2}{3}-2\sqrt{\epsilon})q\\
\displaystyle \qquad\quad\ =-\epsilon+(\frac{1}{3}+\epsilon-2\sqrt{\epsilon})q\geqslant0,
\end{array}
\end{eqnarray*}
and we obtain $q\geqslant\frac{\epsilon}{\frac{1}{3}+\epsilon-2\sqrt{\epsilon}}=\frac{3\epsilon}{1+3\epsilon-6\sqrt{\varepsilon}}.$
It is clear that in order to satisfy above two inequalities, we must ensure that $q\geqslant \{\frac{3\epsilon}{4+3\epsilon},$ $ \frac{3\epsilon}{1+3\epsilon-6\sqrt{\varepsilon}}\}_{max}=\frac{3\epsilon}{1+3\epsilon-6\sqrt{\varepsilon}}.$
Given that $\epsilon\geqslant\frac{2}{3(1-q)}$, we have $\xi_1=1-(1-q)\epsilon\leqslant \frac{1}{3}$ ($\xi_2\leqslant \frac{1}{3}$, $\xi_3\leqslant \frac{1}{3}$), where $\epsilon\in[0.035,0.384]$ is computed by using Matlab (The detailed calculation process is attached in Appendix C).

If the probability of passing the test is high \cite{Winter99,Wilde13}, $P\geqslant1-\epsilon$, then state $\rho$ is ``close" to a certain stabilized state $\rho'$ in the sense of
$$\frac{1}{2}||\rho-\rho'||_1\leqslant 2\sqrt{\epsilon}.$$
Therefore, we prove the soundness.$\square$

\vspace{4mm}
\section{Conclusions}
\label{sec:con}
\vspace{-2mm}
In this section, we first make comparisons with other works \cite{Hayashi18,10Broadbent09,Moe16,7Morimae2014,19Morimae13,20Li14} and then conclude this paper.

In \cite{Hayashi18}, the client is classical and there are two servers labeled prover 1 and prover 2, in which prover 1 prepares the initial state and prover 2 measures the state. In their scheme (Section V.B), two provers are not allowed to communicate once the protocol start which is not practical. However, this situation will not happen in our protocol since we only need a classical verifier and a quantum prover.

In \cite{10Broadbent09}, the server needs to prepare the brickwork state which is difficult in experiment. But it can utilize our protocol model, which can be decribed as follows. First, the unit cluster state is prepared and measured. After that, the server prepares next unit cluster state to measure and the similar process can be repeated until completing the computation. Our protocol model can be used to solve many similar questions. For example in \cite{Moe16,7Morimae2014}, the server can adopt our proposed model to complete a complex quantum computation in experiment.

In \cite{19Morimae13,20Li14}, they directly use the BFK protocol \cite{10Broadbent09} to realize the quantum computation. However, we propose a novel single client-server verifiable blind computation protocol with a new graph state.

To conclude, this article presents a universal measurement-based BQC protocol, which only needs a client and a server. We construct an entangled state with less qubits called latticed state consisted of six-qubit cluster states and eight-qubit cluster states, where the former is mainly used to realize gates S, T, X, Y, Z, I and the latter is mainly applied to obtain gates H, CNOT. Moreover, Alice randomly inserts optimal number of trap qubits to verify the correctness of Bob's measurement outcomes during computing and testing process. Finally, we analyze and prove the correctness, universality, verifiability as well as the blindness of the latticed state, quantum inputs, quantum algorithms and classical outputs. Compared with  the brickwork state, our proposed latticed state is composed of less qubits in the case of realizing a specific quantum computing.

\begin{acknowledgments}
This work was supported by the National Natural Science Foundation of China (Grant No. 62005321).
\end{acknowledgments}

\section*{APPENDIX A}
\begin{figure}[!htp]
  \centering
  \includegraphics[scale=0.5]{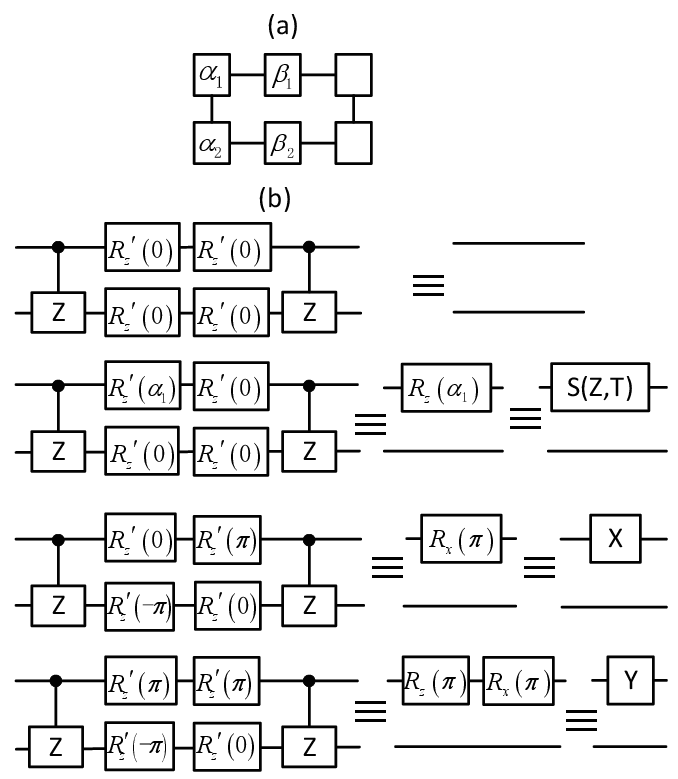}
  \caption{\ Schematic structure of six-qubit cluster states for gates S, T, X, Z, Y, I.}\label{fig3}
\end{figure}

\begin{figure}[!htp]
  \centering
  \includegraphics[scale=0.6]{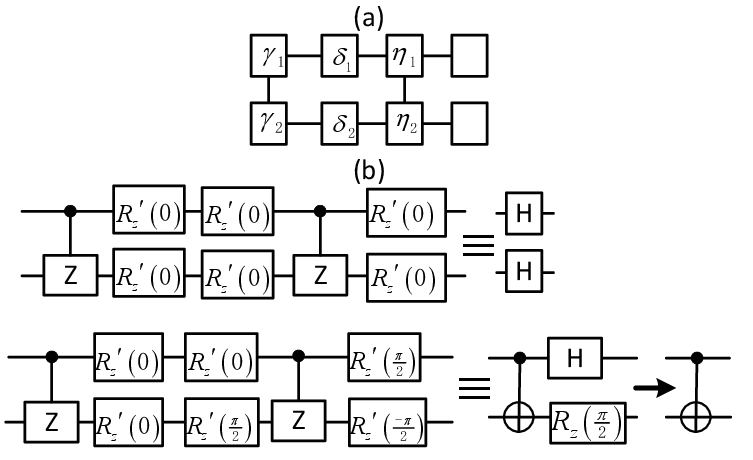}
  \caption{\ Schematic structure of eight-qubit cluster states for gates H and CNOT.}\label{fig4}
\end{figure}

\begin{figure}[!htp]
  \centering
  \includegraphics[scale=0.55]{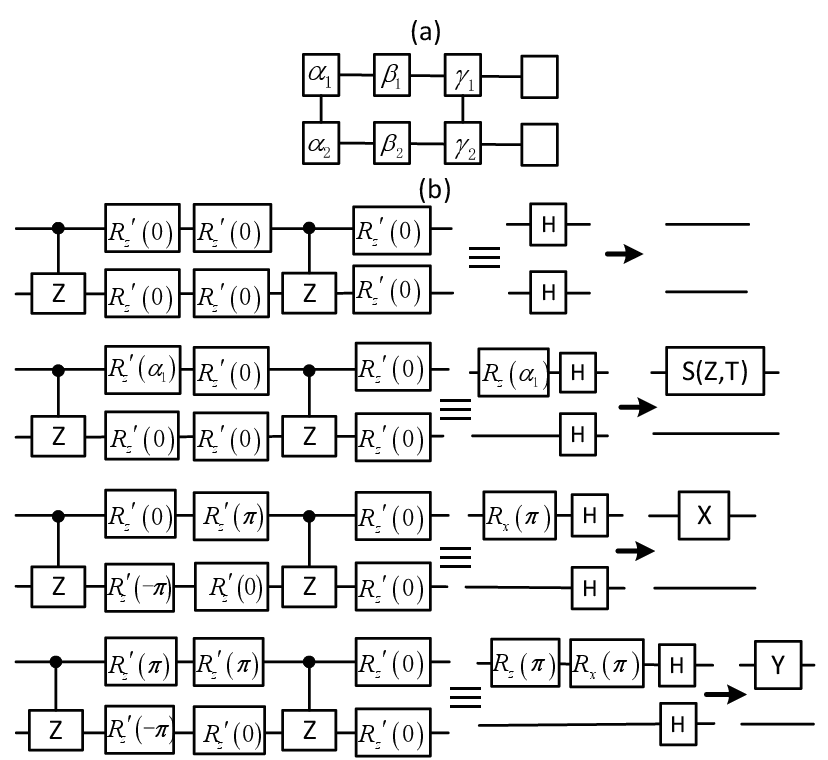}
  \caption{\ Schematic diagram of eight-qubit cluster states for gates H, S, T, X, Y, Z, I.}\label{fig5}
\end{figure}

\begin{figure}[!htp]
  \begin{minipage}[t]{1\linewidth}
  \centering
  \includegraphics[scale=0.27]{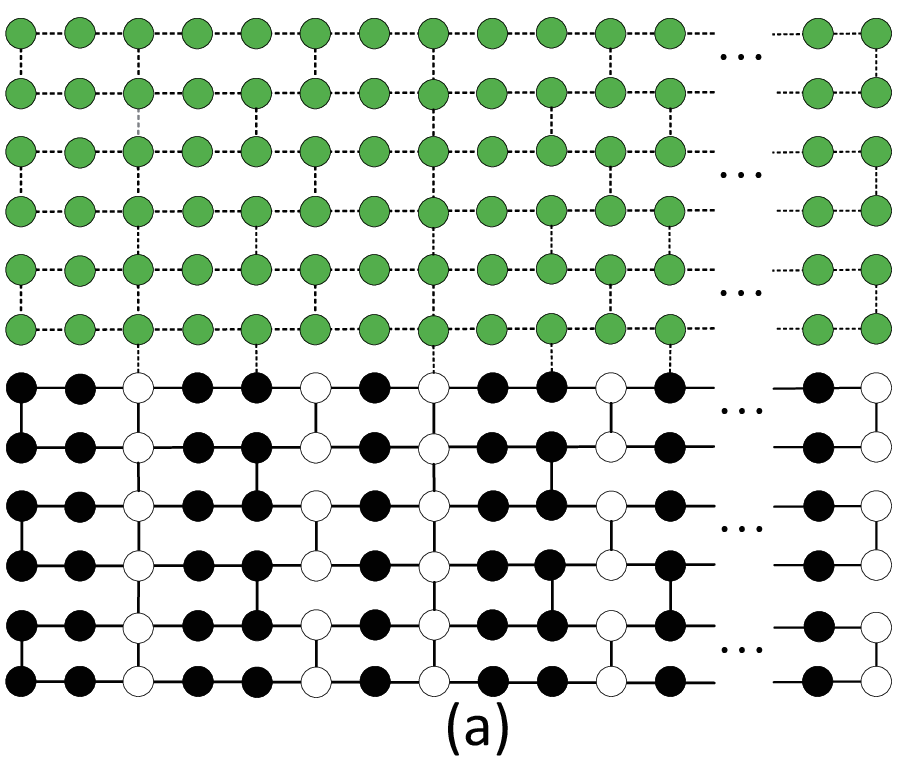}\ \
  \includegraphics[scale=0.27]{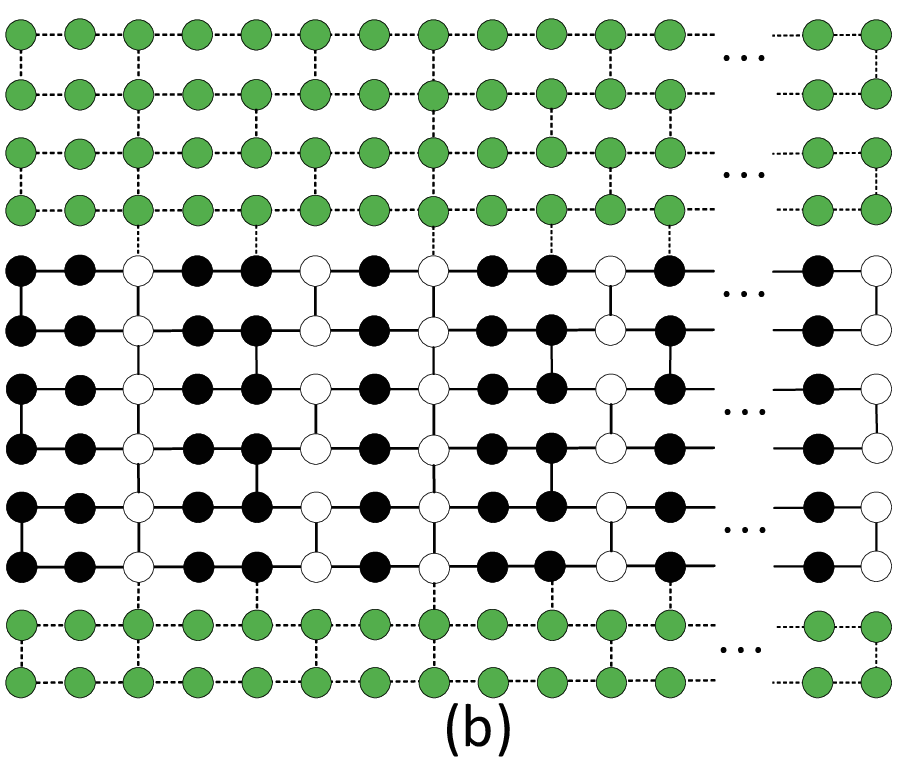}
  \end{minipage}
  \begin{minipage}[t]{1\linewidth}
  \centering
  \includegraphics[scale=0.27]{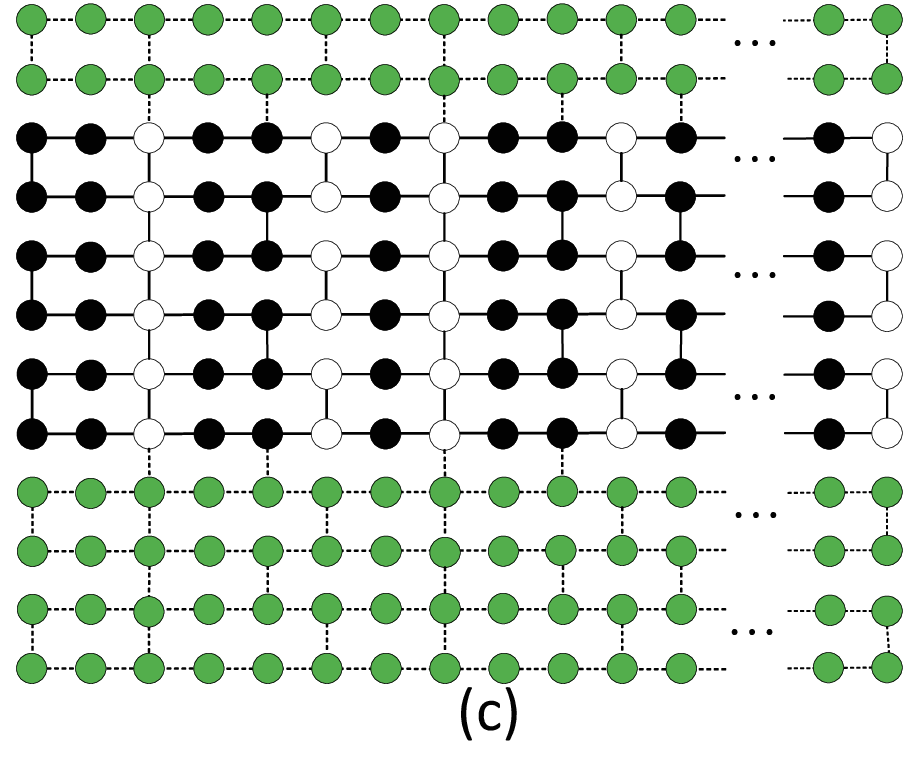}\ \
  \includegraphics[scale=0.27]{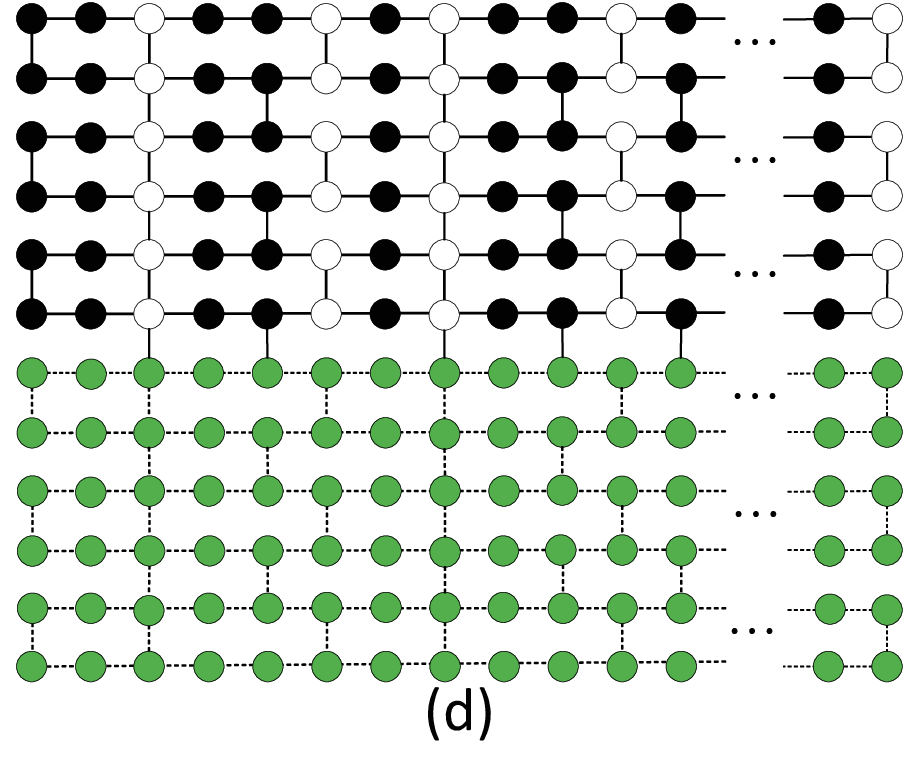}
  \end{minipage}
  \caption{(Color online) Schematic diagram of $|G\rangle$, where qubits connected by the solid lines are entangled but unentangled by the dotted lines.}\label{fig6}
\end{figure}

In this part, we show the schematic structures of six-qubit cluster states and eight-qubit cluster states in Figs. \ref{fig3}, \ref{fig4} and \ref{fig5}. And we also give the form of graph state $|G\rangle$ in Fig. \ref{fig6}.

In Figs. \ref{fig3}, \ref{fig4} and \ref{fig5}, a computation starts with the input information in two left qubits, and measurements are performed from left-to-right. Qubits labelled by $\alpha_x$, $\beta_x$, $\gamma_x$, $\delta_x$, $\eta_x$ ($x=1, 2$) are measured such that the information for each qubit flows to the right along the lines. In general, each horizontal line represents a single qubit propagation, and each vertical connection represents single qubit interaction.

In Fig. \ref{fig3}, $\alpha_x$ and $\beta_x$ are rotation angles in (a). In (b), $R_z(\alpha_x)$ and $R_x(\alpha_x)$ are rotations about the Z axis and X axis, respectively. $R_z'(\theta)=HR_z(\theta)$, it is applicable to Figs. \ref{fig4} and \ref{fig5}. The lines between qubits represent the controlled-Z which are applied before the computation begins.

In Fig. \ref{fig4}, $\gamma_x$, $\delta_x$ and $\eta_x$ are rotations about the X axis and Z axis in (a). Note that, extra gates H and $R_z(\textnormal{-}\frac{\pi}{2})$ need to be performed on the above qubit and the below qubit respectively to get a gate CNOT. If the cluster state does not contain the final quantum outputs, the operation $R_z(\textnormal{-}\frac{\pi}{2})$ will be naturally corrected by performing projective measurements $|\pm_{{\eta_t}-\frac{\pi}{2}}\rangle\Leftrightarrow\frac{e^{\frac{i\pi}{4}}}{\sqrt{2}}[|0\rangle\pm e^{i({\eta_t}-\frac{\pi}{2})}|1\rangle]$.

In Fig. \ref{fig5}, eight-qubits cluster states can also be used to realize single-qubit gates H, S, T, X, Y, Z, I. In this case, Bob needs to perform an undesirable correction operation H on qubits belong to states $|LA\rangle$ or traps $|R_1\rangle$, $|R_2\rangle$. It is obvious that this increases Bob's workload and complexity of this protocol. Therefore, we do not use the eight-qubit cluster states to implement gates H, S, T, X, Y, Z, I as far as possible. In the following, we give the structure of graph state $|G\rangle$ (Fig. \ref{fig6}).

In Fig. \ref{fig6}, the green circles are trap qubits $|R_1\rangle$ and these traps are randomly attached to the latticed state with a certain rule. In such case, Bob cannot precisely extract the traps from state $|G\rangle$, so he learns nothing about the true dimension of the latticed state and the positions of the latticed state.

\section*{APPENDIX B}
The proofs of correctness of X, Y, CNOT are shown in the following.

\emph{Proof:}
We first give the decompositions of gates X, Y, CNOT in Eq.(3).
\begin{eqnarray}
\begin{array}{l}
\displaystyle X=e^{i\frac{\pi}{2}}HR_z(\pi)H R_z(0),\ Y=e^{\textnormal{-}\frac{i\pi}{2}}HR_z(\pi)H R_z(\pi),\\
\displaystyle CNOT=(R_z(\frac{\pi}{2})\otimes R_x(\frac{\pi}{2}))CZ(I\otimes R_x(\textnormal{-}\frac{\pi}{2}))CZ.
\end{array}
\end{eqnarray}
where $(R_x(\pi)\otimes I)CZ=e^{\frac{i\pi}{2}}CZ(R_x(\pi)\otimes R_z(\pi))$ and $(Y\otimes I)CZ=CZ(Y\otimes Z)$.

\begin{figure}[!htp]
  \centering
  \includegraphics[scale=0.6]{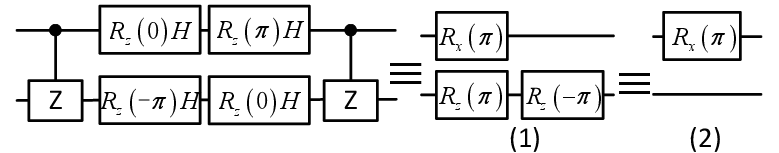}
  \caption{Simplified process of gate X.}\label{fig10}
\end{figure}

Fig. \ref{fig10} gives out the simplified process of gate X. For the above qubit, we have $HR_z(\pi)H R_z(0)=R_x(\pi)$. According to the equation $(R_x(\pi)\otimes I)CZ=e^{\frac{i\pi}{2}}CZ(R_x(\pi)\otimes R_z(\pi))$, we can move the $R_x(\pi)$ from the right of the first $CZ$ to the left with auxiliary gate $R_z(\pi)$ on the below qubit. We set the angles $\alpha_2=\textnormal{-}\pi$, $\beta_2=0$ and use the equation $(R_z(\theta)\otimes I)CZ=CZ(R_z(\theta)\otimes I)$ to eliminate the influence of $R_z(\pi)$ to get the circuit (1). Finally, we realize gate $X$ on the above qubit, so does the below qubit.

\begin{figure}[!htp]
\begin{minipage}[t]{0.5\linewidth}
\centering
\includegraphics[width=2.0in]{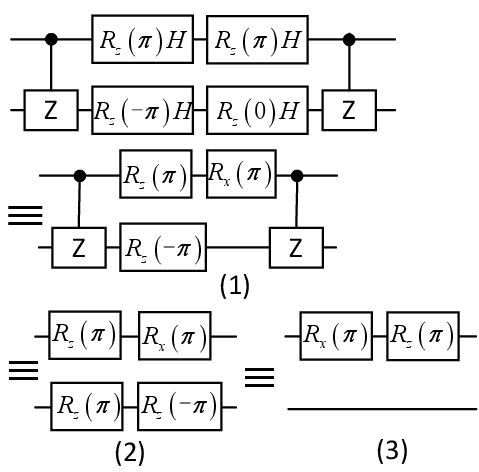}
\caption{\ Simplified process of gate Y.}\label{fig11}
\label{frame}
\end{minipage}%
\begin{minipage}[t]{0.5\linewidth}
\centering
\includegraphics[width=1.8in]{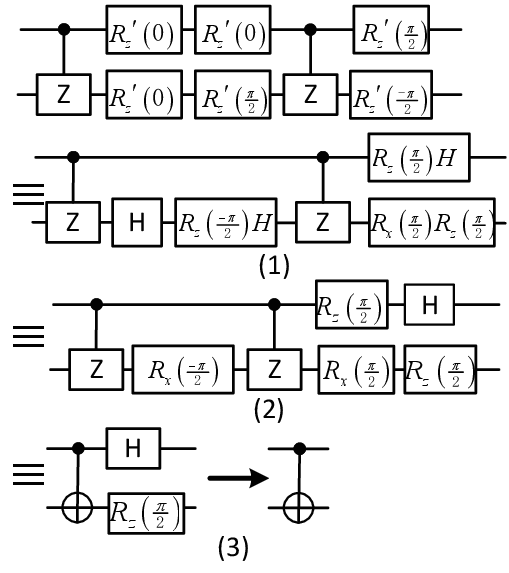}
\caption{\ Simplified process of gate CNOT.}\label{fig12}
\label{label}
\end{minipage}
\end{figure}

The simplified process of gate Y can be seen in Fig. \ref{fig11}. By the relationship $HR_z(\pi)H=R_x(\pi)$ in the above line, we get the circuit (1). Similar to gate X, we get the circuit (2) according to the equations of $(R_x(\pi)\otimes I)CZ=e^{\frac{i\pi}{2}}CZ(R_x(\pi)\otimes R_z(\pi))$ and $(R_z(\pi)\otimes I)CZ=CZ(R_z(\pi)\otimes I)$. Therefore, we realize gate $Y$ on the above qubit, so does the below qubit.

The simplified process of gate CNOT can be seen in Fig. \ref{fig12}. Through the relationship $HR_z(0)H$$R_z(0)=I$, the above line is I gate so we get the circuit (1). By the relationship $HR_z(\textnormal{-}\frac{\pi}{2})H=R_x(\textnormal{-}\frac{\pi}{2})$ and $R_z(\textnormal{-}\frac{\pi}{2})H=R_x(\frac{\pi}{2})R_z(\frac{\pi}{2})$, we get the circuit (2). Via the relationship $(R_z(\frac{\pi}{2})\otimes R_x(\frac{\pi}{2}))CZ(I\otimes R_x(\textnormal{-}\frac{\pi}{2}))CZ=CNOT$, we get the gate CNOT after correcting H and $R_z(\textnormal{-}\frac{\pi}{2})$.$\square$

\section*{APPENDIX C}
Here, we give a detailed calculation process for the range of $\epsilon$. Since $\epsilon\geqslant\frac{2}{3}\cdot\frac{1}{1-q}\geqslant\frac{2}{3}\cdot\frac{1+3\epsilon-6\sqrt{\epsilon}}{1-6\sqrt{\epsilon}}$, we get $18\epsilon\sqrt{\epsilon}+3\epsilon-12\sqrt{\epsilon}+2\leqslant0$. Suppose a function $f(x)=18x\sqrt{x}+3x-12\sqrt{x}+2$, the first-order derivative is $f'(x)=27\sqrt{x}+3-\frac{6}{\sqrt{x}}$. When $f'(x)$ equals to $0$, the solution is $x\approx0.175$ so we obtain $f(0.175)=-1.1772$. The second-order derivative of $f(x)$ is $f^{''}(x)=\frac{27}{2\sqrt{x}}+\frac{3}{x\sqrt{x}}$, and we can know $f^{''}(0.175)>0$. According to the sufficient conditions of extreme value, $f(0.175)=-1.1772$ is the minimum value. When $f(x)=0$, we get $x_1\approx0.035, x_2\approx0.384$ calculated by Matlab. By analyzing the relationship of $x, f'(x)$ and $f(x)$, we get the conclusion: the function $f(x)$ is decreasing when $x\in[0,0.175)$, while it is increasing when $x\in(0.175,1]$. It is easy to get $x\in[0.035, 0.384]$ when $f(x)\leqslant0$. Therefore, the range of $\epsilon$ is $[0.035, 0.384].$

Moreover, for $q\geqslant\frac{3\epsilon}{1+3\epsilon-6\sqrt{\varepsilon}}$, we verify that the range of $q$ is $[0,1]$. Suppose a function $g(y)=\frac{3y}{1+3y-6\sqrt{y}}$, we compute the first-order derivative $g'(y)=\frac{3-9\sqrt{y}}{(1+3y-6\sqrt{y})^2}$. The function $g(y)$ is increasing if $g'(y)=\frac{3-9\sqrt{y}}{(1+3y-6\sqrt{y})^2}\geqslant0$ with $y\in[0,\frac{1}{9}]$. Otherwise, $g(y)$ is decreasing with $y> \frac{1}{9}$. Naturally, we obtain $g(y)_{max}=g(\frac{1}{9})=-\frac{1}{2}$. For $\epsilon\in[0.035, 0.384]$, it is obviously that $\frac{1}{9}\in[0.035, 0.384].$ Hence, it is reasonable for $q\in[0,1].$

\end{document}